\documentclass{epl}

\title{Elasticity of an interfacial particle raft }
\shorttitle{Particle rafts}
\author{D. Vella\inst{1} \and P. Aussillous\inst{2} \and L. Mahadevan\inst{1} \footnote{Email: lm@deas.harvard.edu}}
\institute{
  \inst{1}  Division of Engineering and Applied Sciences, Harvard University, Cambridge, MA 02138, USA \\
  \inst{2}   Department of Applied Mathematics and Theoretical Physics, 
University of Cambridge, Wilberforce Road, Cambridge, CB3 0WA,  UK
}

\begin{document}

\maketitle

\begin{abstract}
We study the collective behaviour of a close packed monolayer of non-Brownian particles at a fluid-liquid interface. Such a particle raft forms a two-dimensional elastic solid and can support anisotropic stresses and strains, e.g.  it buckles in  {uniaxial} compression and cracks in tension. We characterise this solid in terms of a Young's modulus and Poisson ratio derived from simple theoretical considerations and show the validity of these estimates by using an experimental buckling assay to deduce the Young's modulus.
\end{abstract}

\section{Introduction}

Particle covered liquid interfaces are increasingly being exploited in a wide variety of technological and medical applications \cite{Binks}. Coating a liquid drop with a hydrophobic powder renders the drop non-wetting with the resulting liquid marbles free to roll on rigid surfaces  or even float on water \cite{Aussillous}, a feature that has also proven useful as an adaptation to life on small scales \cite{Richard}. In a similar vein, it is hoped that by encapsulating the active ingredients of drugs within a monolayer of colloidal particles (thereby forming a \textit{colloidosome} \cite{Dinsmore}) more medicines will soon be administered by inhalation, so improving their efficacy. In the latter case, the particle coated interface exists only in the preliminary stages of manufacture, but in neither case have previous investigations been concerned with understanding the properties of particle monolayers at liquid interfaces. In this article, we show that in fact such particle monolayers behave collectively like 2-dimensional elastic solids and further, we characterize the properties of this two-dimensional solid.

\begin{figure}
\centering
\includegraphics[height=4cm]{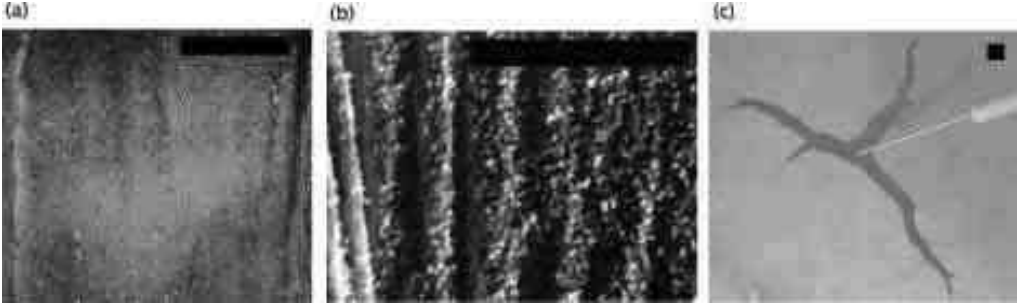}
\caption{Typical wrinkling patterns observed experimentally: (a) Polysytrene particles with mean diameter $300$\un{\mu m} and (b) Lycopodium particles with mean diameter $30$ \un{\mu m}. (c) shows the crack in a monolayer of particles with diameter in the range $150-212$ \un{\mu m} near equilibrium. Here the needle is used only to place surfactant on the surface; the gradient in interfacial tension causes a crack to open. The scale bar in each image corresponds to $5$ \un{mm}.}
\label{pics}
\end{figure}

A variety of solid-like behaviours are observed for monolayers of hydrophobic particles sprinkled densely onto an air-water interface for a wide range ($2.5 $ \un{\mu m} - $6$ \un{mm}) of particle sizes. For example, such a monolayer buckles under sufficient static compressive loading  (see figs. \ref{pics} (a) and (b)) demonstrating that it can support an anisotropic stress. This stress state can only be supported by a material with a non-zero shear modulus, which is the signature of a solid. Once the compressive stress is removed, the monolayer returns rapidly ($\sim {\cal O}(0.1)$s) to the undeformed state, ironing out the wrinkles formed by the buckling. This elasticity is also reminiscent of a solid and is in stark contrast to what is commonly observed in both dry and wet granular systems, which also resist compression but do not, on the whole, return to the undeformed state once the compressive stress is removed. 

Furthermore, the solid will also fracture under relatively small tensions and is manifested by the spectacular branching fracture induced by the addition of a drop of surfactant as shown in fig. \ref{pics} (c). Here, the disparity in surface tension  between the clean and contaminated areas is enough to open up a tensile crack that propagates with surfactant advected to the crack tip by Marangoni flows. 

These behaviours are curious since the monolayer has solid-like properties only in the presence of the liquid and is reduced to a cohesionless powder once this liquid is evaporated naturally, making this an unusual composite material. Capillary forces are responsible for the formation of the solid monolayer via the aggregation of particles trapped at the interface and give the monolayer cohesion under deformation.  The particle monolayers we study are distinct from other two-dimensional interfacial systems such as Langmuir monolayers \cite{Gaines} because the large size of the particles (diameters of up to 6 \un{mm}) makes them non-Brownian objects which  interact solely via capillary attraction. In this respect they are similar to conventional bubble rafts \cite{Bragg}, and so we dub them particle rafts. However, bubble rafts of a similar scale are more easily deformable in compression  and stack into multiple layers under the effects of  large compressive stresses rather than forming larger scale wrinkles. This latter difference is due to the fact that bubbles are considerably less dense than the bulk fluid, ensuring that it is energetically cheaper to stack them than to deform the bulk interface significantly, which is a requirement of wrinkle formation. For a monolayer particle raft loaded in  {uniaxial} compression, the  steric interaction between particles allows for anisotropic stresses while {attractive} capillary forces prevent stress relaxation due to particle creep in the direction orthogonal to the {uniaxial} compression. 

Although the dynamics of crack propagation in this system are of interest, we defer further discussion to the future and focus in this article on characterising the elastic properties of the particle raft. In this system, it seems reasonable to assume that the particle raft behaves approximately as an isotropic solid, in which case its elastic behaviour can be completely characterised in terms of the Young's modulus, $E$, and Poisson ratio, $\nu$; in the next section, we shall determine these parameters theoretically. 

\section{A Simple Model}

Microscopic theories of elastic solids usually rely on the existence of a minimum in the interaction potential of the constituent particles, so that the elastic modulus is related to the curvature of this potential at the minimum. In the present situation, however, the following modification of a simple argument by Lucassen \cite{Lucassen} suffices.

For a conventional two-dimensional elastic solid, the mean stress, $\bar{\sigma}= \frac{\sigma_1+\sigma_2}{2}$, and strain, $\bar{\epsilon}=\frac{\epsilon_1+\epsilon_2}{2}$ (where the subscripts $1,2$ denote the two principal directions), are related by \cite{Landau}:
\begin{equation}
\bar{\epsilon}=\frac{1-\nu}{E}\bar{\sigma}
\label{2deq}
\end{equation} where $\nu$ is the Poisson ratio of the solid. If we replace the stress by a thickness-averaged isotropic tension, $\tau\equiv\bar{\sigma} d$, then this relation reads:
\begin{equation}
\bar{\epsilon}=\frac{1-\nu}{Ed}\tau
\end{equation} We then have:
\begin{equation}
\frac{1-\nu}{Ed}=\frac{d\bar{\epsilon}}{d\tau}=\frac{1}{A}\frac{dA}{d \tau}=\frac{1}{A_l+A_s}\frac{d(A_l+A_s)}{d \tau}
\label{average}
\end{equation} where $A_l$ is the area of the system covered by liquid, $A_s$ is the area covered by solid particles and $A=A_l+A_s$ is the total area of the system. The area covered by particles is constant regardless of the tension applied since  the solid particles are effectively rigid at the individual level in comparison to the raft as a whole which collectively can be soft. In addition, $A_l\frac{d\tau}{dA_l}\propto \gamma$ (with a constant of proportionality that may, in general, be different to $1$) so that (\ref{average}) can be rearranged to give the effective Young's modulus: 
\begin{equation}
E\propto \frac{1-\nu}{1-\phi}\frac{\gamma}{d}
\label{law}
\end{equation} where $\phi=A_s/A$ is the solid fraction of the interface. 

While naive dimensional analysis would predict that $E\sim \gamma/d$, by recalling (\ref{2deq}), we might expect that $E$ should only enter through the combination $E/(1-\nu)$. This would lead us to expect that $E\sim (1-\nu)f(\phi)\gamma/d$ for some function $f(\phi)$, which we know must have the feature that as $\phi\rightarrow 1$, $f\rightarrow\infty$, since the particles are assumed rigid. The argument leading to (\ref{law}) showed that $f(\phi)\propto(1-\phi)^{-1}$, which does have the desired divergence properties and is perhaps the most natural such function.

\begin{figure}
\centering
\includegraphics[width=8cm]{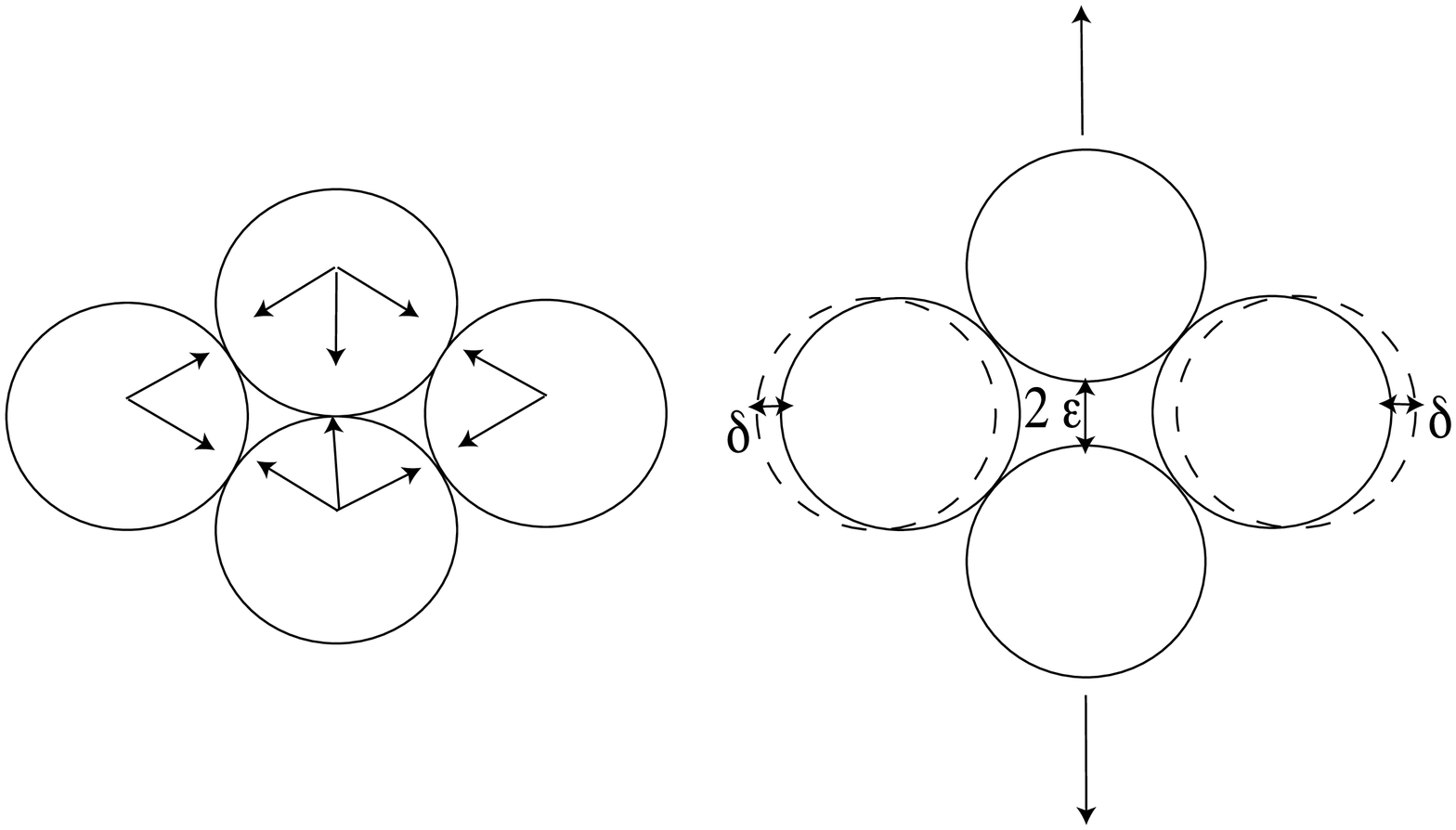}
\caption{Diagram showing the Poisson effect for four hexagonally close packed spheres (forming a rhombic cell) interacting via a purely central force.}
\label{poisson}
\end{figure}

 {An estimate of} the value of $\nu$ may be determined from the assumption that the particles are hexagonally close packed discs that interact only by means of an  {attractive} central force  {and a repulsive steric interaction}: a reasonable approximation for the case of spherical particles. Figure \ref{poisson} demonstrates the origin of the Poisson effect for a single rhombic cell in the lattice in which the two central particles are displaced by a distance $2\epsilon$. From elementary geometry, the rhombus joining the centres of the four particles in the undeformed state has width $2\sqrt{3}R$ and height $2R$, where $R$ is the particle radius. During the deformation, the width decreases to $2\sqrt{3}R\sqrt{1-2\epsilon/3R}\approx 2\sqrt{3}R(1-\epsilon/3R)$ and the height increases to $2(\epsilon+R)$ so that the Poisson ratio is simply $\nu=1/\sqrt{3}$.

Under the same assumption that the particles have a perfectly circular cross-section and are monodisperse, we may determine from elementary geometry that in hexagonal close packing, $\phi = \pi/2\sqrt{3}$. This gives an estimate of the typical value of the Young's modulus:
\begin{equation}
E\sim 2\frac{\sqrt{3}-1}{2\sqrt{3}-\pi}\frac{\gamma}{d}\approx 4.54 \frac{\gamma}{d}
\label{high}
\end{equation} 

 {The calculations above are based on the assumption that the particles are hexagonally close packed discs when, in fact, their shape is irregular leading to non-ordered packings. However, this approximation provides estimates for the values of $\nu$ and $\phi$, which can then be used to obtain an estimate for the prefactor in (\ref{high})}

The results (\ref{law},\ref{high})  show that the elastic properties of the system  are determined just by the surface tension, $\gamma$, and the particle diameter, $d$. In particular, there is no dependence on  {the} particle elasticity {since we have neglected the possibility of particle deformations in setting $\frac{dA_s}{d\tau}=0$ in the argument leading to (\ref{law}). This is consistent with the picture that it is much easier to deform the layer by collective relative motion of particles than to deform the individual particles. We also see that the shape of the particles enters only via the dimensionless parameters $\phi$ and $\nu$.}

Physically, the argument above shows that the elasticity of the particle raft arises because of changes in the interfacial area that are a result of the imposed deformation. In the undeformed state, the interfacial area is minimized (since this state is an equilibrium state) subject to the constraint that the solid particles are not deformable. Once a deformation occurs, the particles rearrange themselves to accommodate this imposition but the interfacial area associated with the new state is necessarily larger than that of the equilibrium state (since we have more constraints under deformation) and so energy is stored in the additional interface created. 

\section{Experiment: a buckling assay}

\begin{figure}
\centering
\includegraphics[height=4cm]{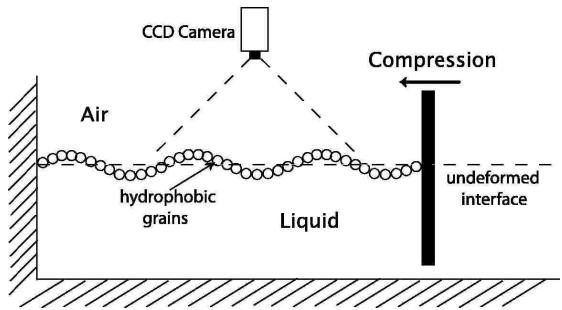}
\caption{Schematic of the experiment used to observe the buckling instability pictured in fig. \ref{pics}.}
\label{setup}
\end{figure}

To test the simple expression in (\ref{law}), we used an elastic buckling assay of the particle raft to determine the wavelength (see fig. \ref{pics}) of the instability and thence infer the value of the Young's modulus as a function of the particle size.   To measure this dependence, each powder was in turn deposited  by sprinkling onto an air-liquid interface contained in a home-made Langmuir trough (dimensions $11$cm$\times 5$cm$\times$1cm  {for small particles sizes, $38.5$cm$\times 18.5$cm$\times4.2$cm for large particles}). The barrier of the Langmuir trough consisted of a slide of the same width as the trough itself but slightly smaller thickness allowing the level of the liquid to equilibrate by flow beneath the barrier (see fig. \ref{setup}). The barrier was used to compress the particle monolayer until the buckling instability sets in; a schematic of the experimental setup is shown in fig. \ref{setup}. The rate of compression leading to the buckling instability is not a factor in determining the wavelength of wrinkles produced as we measure the wavelength in the static configuration once any transients due to the induced flow in the liquid sublayer have decayed.   {Photographs of the wrinkled surface were taken using a high resolution CCD camera (1600$\times$1200 pixels) for different runs. Single wavelengths were measured at different points of different photos  (at least ten measurements) using image analysis software (ImageJ, NIH) and the mean value calculated.}

The particles used were primarily made from Pliolite (Eliokem), a highly hydrophobic material, ground and sorted into different  samples ranging in mean diameter from $40$ \un{\mu m} to $2$mm.  However, we also used a number of other hydrophobic powders both to obtain a wider size range and   test the dependence of the results on the particles used. These included silanised Lycopodium \cite{Aussillous} (grain size $\approx 20$ \un{\mu m}), graphite powder (grain size $\approx 2.5$\un{\mu m}), powdered Chemigum (Eliokem, grain size $\approx 150$ \un{\mu m}) and polystyrene powder of different mean grain sizes (ranging from $300$ \un{\mu m} to $7$ \un{mm}) - the formation of a two-dimensional solid apparently only requires that these particles be trapped at the liquid-air interface. For particles larger than $40 $ \un{\mu m}, we were able to use the CCD camera to measure the particle size and comparing 20 grains of a particular sample gave us estimates for the polydispersity of that sample. For samples below this threshold, we relied on the size ranges specified by the manufacturer.

Since the powders used were neither monodisperse nor uniformly spherical, the solid fraction of the monolayer, $\phi$, was, in general, different from the maximum possible packing density of two-dimensional discs (which is close to $91\%$). Therefore, $\phi$ was measured just prior to the onset of instability using image analysis software (ImageJ, NIH) by changing the threshold on each image (the good contrast between liquid and particle phases ensuring that the exact value of the threshold used is immaterial). The experimental measurements of $\phi$ gave a mean value $\bar{\phi}=0.85$ with standard deviation $0.06$.

Most experiments were performed with the base liquid being water ($\rho=1000$\un{kg m^{-3}} and $\gamma=72$\un{mN m^{-1}}) but we also performed experiments with saline solution ($5\%$, by weight, \chem{NaCl} so that $\rho=1034$\un{kg m^{-3}} and $\gamma=74$\un{mN m^{-1}}) to test for the possibility of electrostatic interactions mediated by the water. Previously it has been shown that the presence of the powder does not alter the surface tension coefficient as determined from comparisons between experiments on liquid marbles and numerical results for pure non-wetting drops \cite{Aussillous2} and so in the subsequent analysis we used surface tension coefficients for pure liquid samples. 

\begin{figure}
\centering
\includegraphics[width=6.9cm]{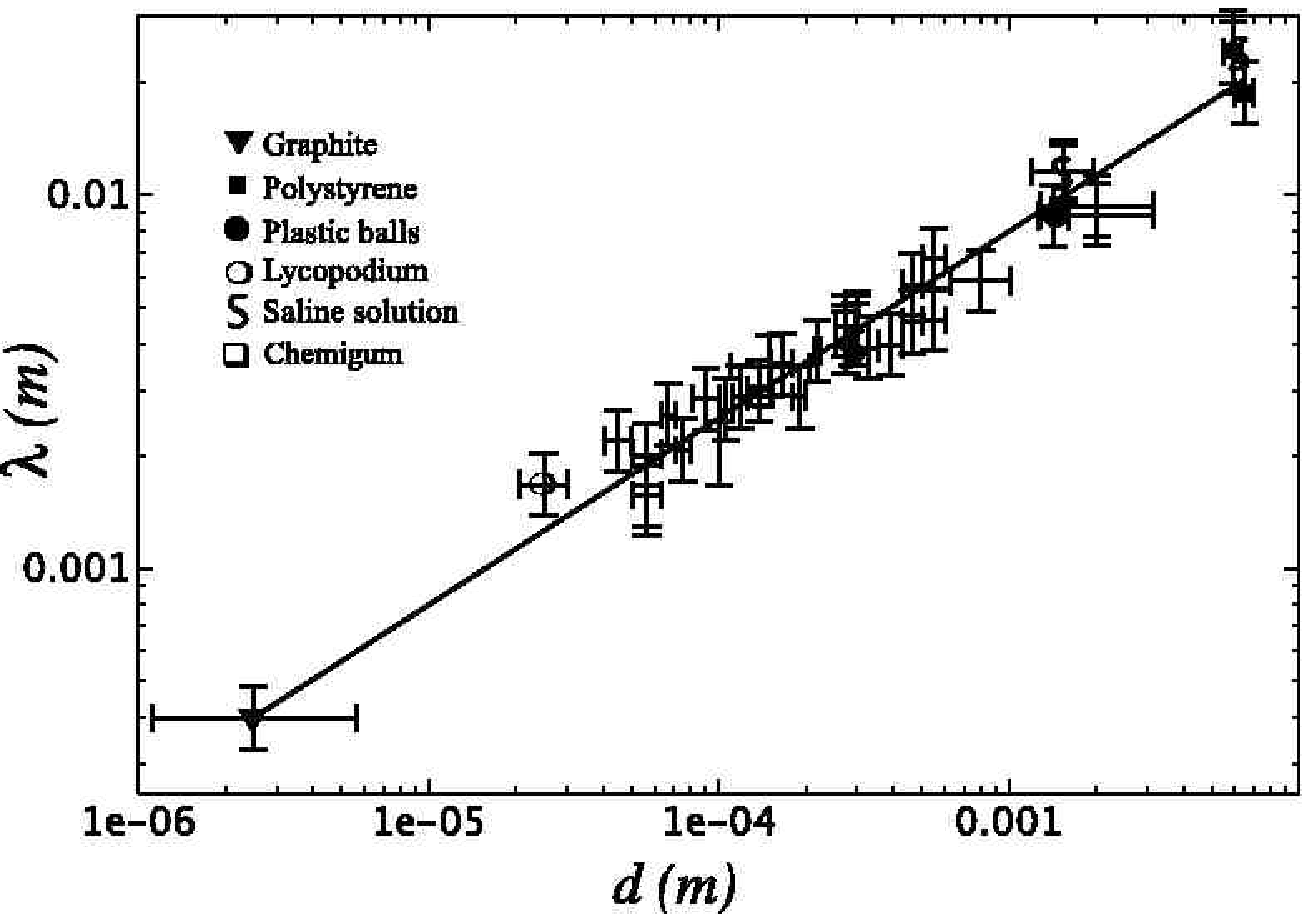}
\includegraphics[width=7.2cm]{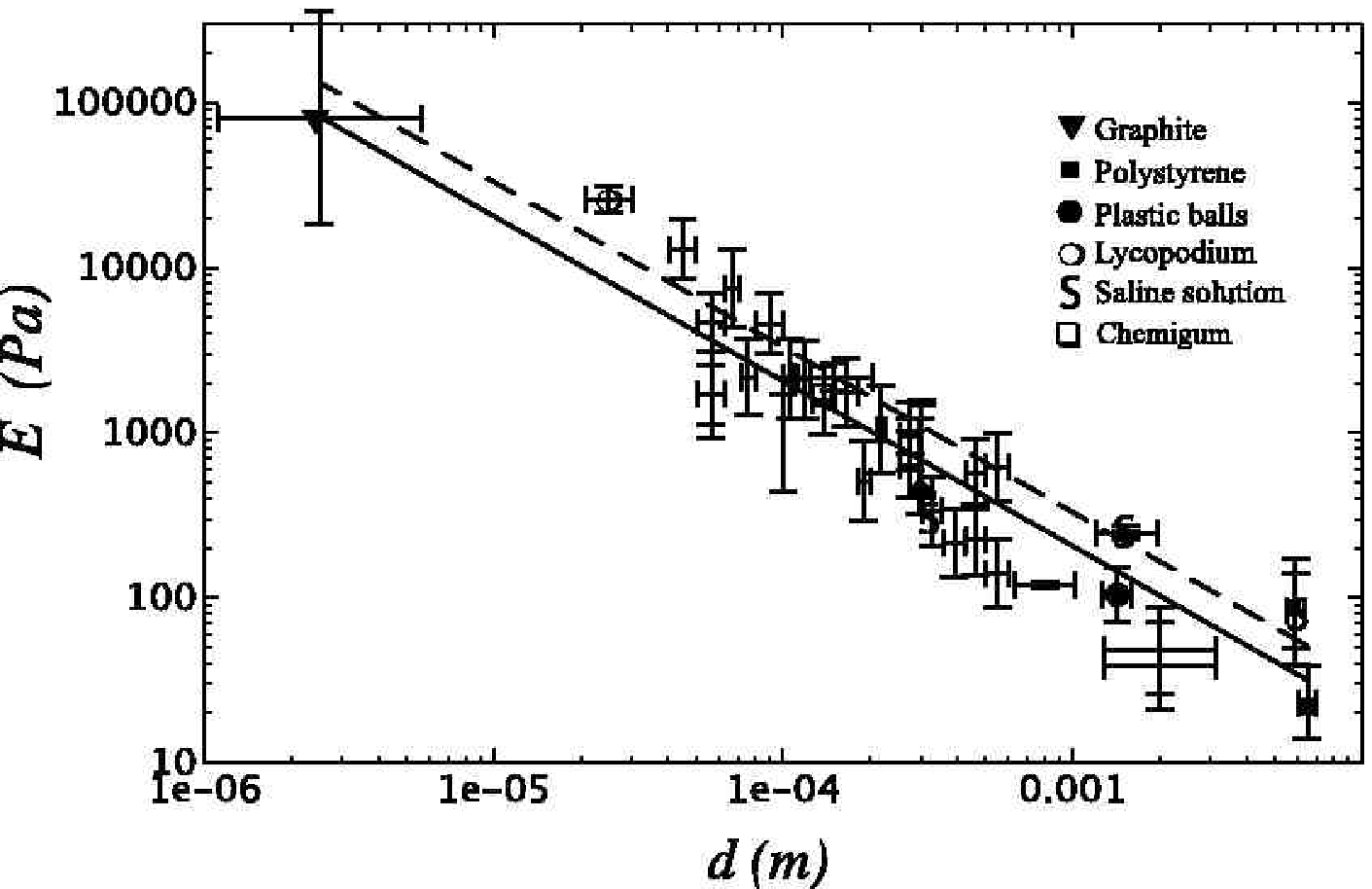}
\caption{Experimental measurements of the dependence of the ripple wavelength $\lambda$ on the particle diameter, $d$.  All experimental points are for pliolite at an air-water interface unless otherwise specified by the key. The solid line shows the theoretical prediction (\ref{lampred}). The error bars in the X-direction  indicate the polydispersity of the sample, while the error bars in the Y-direction indicate the polydispersity in the measured wavelength.}
\label{lamdep}
\caption{The Young's modulus as a function of the particle diameter for a range of particles. The solid line shows the theoretical relation $E= 2.82 \gamma/d$. The dashed line shows the prediction from (\ref{high}). All data points are for pliolite at an air-water interface unless indicated otherwise.}
\label{youngmod}
\end{figure}

In fig. \ref{lamdep}, we plot the wavelength as a function of the mean particle diameter for a range of different particle sizes and the two different base liquids. The results of these experiments lie on a single line and so we conclude that the precise geometry of the particles used as well as electrostatic effects are of minimal significance \cite{note}. 

\section{Determination of $E$ from Buckling}

Classical elasticity theory \cite{Landau}  allows us to deduce the effective mechanical properties of this unusual two-dimensional solid from the measured wavelength which is assumed to be large compared to the particle size. We assume that the raft behaves as a thin sheet of isotropic, homogeneous elastic material of thickness equal to the particle diameter $d$ in plane stress.  At the onset of the buckling instability, the  deflection of the raft is small and so its vertical displacement, $h(x)$, measured relative to its equilibrium position in the absence of any compressive force satisfies the beam equation \cite{Landau}:
\begin{equation}
B\frac{\partial^4 h}{\partial x^4}+T\frac{\partial^2 h}{\partial x^2}+\rho g h=0
\label{eqm}
\end{equation} where $B=Ed^3/12(1-\nu^2)$ is the bending stiffness of the sheet, $T$ is the compressive force per unit length in the sheet, $\rho$ is the density of the liquid whose weight provides the restoring force, and $g$ is the acceleration due to gravity. Substituting $h(x) = A \sin (2 \pi x/\lambda)$ into (\ref{eqm}) gives us an expression for the compressive force $T$ as a function of $\lambda$; minimizing the result yields the selected wavelength $\lambda =\pi (4 Ed^3/(3 \rho g(1-\nu^2)))^{1/4}$. Rewriting this expression gives the Young's modulus $E$ in terms of the observed wavelength and the particle diameter as:

\begin{equation}
E=\frac{3}{4\pi^4}\frac{\rho g (1-\nu^2) \lambda^4}{d^3}
\label{young}
\end{equation}
Using (\ref{young}) the data shown in fig. \ref{lamdep} can be used to calculate the effective Young's modulus of the particle raft as a function of particle diameter $d$, with the result shown in fig. \ref{youngmod}. We see that the Young's modulus is approximately inversely proportional to the particle diameter with a constant of proportionality close to the surface tension coefficient $\gamma$, as is predicted by our simple model (\ref{law}).

Substituting the mean value for solid volume fraction measured experimentally $\phi=0.85$  and $\nu=1/\sqrt{3}$ into (\ref{law}) we get $E=2.82 \gamma/d$, indicated by the solid line drawn in fig. \ref{youngmod}, while the dashed  line corresponds to (\ref{high}). Linear regression performed on this data suggests that $E=0.154d^{-1.03}$ in SI units, which compares favourably to the theoretical prediction based on $\phi=0.85$, which is $E=0.205 d^{-1}$. Finally, substituting the expression for the Young's modulus (\ref{law})  with proportionality constant unity and rearranging (\ref{young}) yields an expression for the wavelength of the wrinkles  :
\begin{equation}
\lambda=\pi\left( \frac{4}{3(1-\phi)(1+\nu)}\right)^{1/4}\sqrt{L_cd}
\label{lampred}
\end{equation} where $L_c=\sqrt{\gamma/\rho g}$ is the capillary length, which result is plotted as the solid line in fig. \ref{lamdep} and fits the experimental data well. Linear regression on the data suggests that $\lambda=0.235 d^{0.49}$, which is close to the theoretically predicted result, which in SI units reads $\lambda=0.253 d^{1/2}$.

\section{Conclusions}

We have seen that a two-dimensional monolayer of packed particles trapped at a liquid interface behaves like a two-dimensional elastic solid whose properties we have characterised. Our theoretical model for this behaviour shows that the properties of the particle raft are relatively insensitive to the details of the capillary interaction but do depend on the geometry of packing and has been verified by experiments on a wide range of particle rafts. 
Some variations in the experimental results could be accounted for by the polydispersity of the particle sizes and shapes and the resulting non-uniform packing. However, in such cases the value of $\nu$ is different from $1/\sqrt{3}$, introducing an additional complexity that is beyond the scope of the present work.

Many interesting questions associated with particle rafts involving both their dynamic and non-linear response remain to be answered. For instance, it is not clear what governs the speed of propagation of the cracks alluded to in the Introduction. Plastic behavior induced by shear in a Couette geometry shows that the particle rafts respond by localization of the deformation to a small region of the sheet, as for two-dimensional bubble rafts  \cite{Lauridsen}. However our system may be more amenable to  investigation since individual grains  in particle rafts are effectively rigid. Indeed, this suggests that particle rafts could serve as a model two-dimensional granular system.

\acknowledgments
D.V. is supported by the Choate Fellowship at Harvard, P.A. is supported by an EU Marie-Curie Fellowship at Cambridge, and L.M. is supported by the US Office of Naval Research  and  the US National Institutes of Health.


\begin{thebibliography}{18}
\bibitem{note} There is the possibility of electrostatic interactions through the air, although this effect should also be well screened in general.
\end{thebibliography}
\end{document}